\documentclass[10pt,journal]{IEEEtran}%
\usepackage{cite}
\usepackage{graphicx}
\usepackage{balance}
\usepackage{epsfig}
\usepackage{epstopdf}
\usepackage{enumitem}

\ifCLASSINFOpdf
\else 
\fi
\usepackage{colortbl}

\begin{document}
\title{Understanding the Computational Requirements of Virtualized Baseband Units using a Programmable Cloud Radio Access Network Testbed}
% \title{Understanding Computational Characteristics \rev{(or Requirements?)} of Virtualized Baseband Units using a Programmable Cloud RAN Testbed}
% \title{Experimental Testbed for Low-Latency Virtualized Cloud Radio Access Networks}
%\title{Implementation and Evaluation of A Low-latency Cloud Radio Access Network Testbed}
\author{{\bf Tuyen X. Tran, Ayman Younis, and Dario Pompili}\\Department of Electrical and Computer Engineering\\
Rutgers University--New Brunswick, NJ, USA\\
E-mails: \{tuyen.tran, a.younis, pompili\}@rutgers.edu
}

\maketitle

\markboth{In Proceedings of the IEEE International Conference on Autonomic Computing (ICAC), July 2017 }
{Shell \MakeLowercase{\textit{et al.}}: Bare Demo of IEEEtran.cls for Computer Society Journals}

\thispagestyle{empty}
\pagenumbering{arabic}
\IEEEdisplaynotcompsoctitleabstractindextext
\IEEEpeerreviewmaketitle

\begin{abstract} %Harry rewrote abstract
Cloud Radio Access Network~(C-RAN) is emerging as a transformative architecture for the next generation of mobile cellular networks. In C-RAN, the Baseband Unit~(BBU) is decoupled from the Base Station~(BS) and consolidated in a centralized processing center. While the potential benefits of C-RAN have been studied extensively from the theoretical perspective, there are only a few works that address the system implementation issues and characterize the computational requirements of the virtualized BBU. In this paper, a programmable C-RAN testbed is presented where the BBU is virtualized using the OpenAirInterface~(OAI) software platform, and the eNodeB and User Equipment~(UEs) are implemented using USRP boards. Extensive experiments have been performed in a FDD downlink LTE emulation system to characterize the performance and computing resource consumption of the BBU under various conditions. It is shown that the processing time and CPU utilization of the BBU increase with the channel resources and with the Modulation and Coding Scheme~(MCS) index, and that the CPU utilization percentage can be well approximated as a linear increasing function of the maximum downlink data rate. These results provide real-world insights into the characteristics of the BBU in terms of computing resource and power consumption, which may serve as inputs for the design of efficient resource-provisioning and allocation strategies in C-RAN systems.
\end{abstract}

\IEEEpeerreviewmaketitle

%---------------------------------------------------   
\section{Introduction}
% Over the last few years, the proliferation of personal mobile-computing devices such as tablets and smart phones, along with a plethora of data-intensive mobile applications, has resulted in a tremendous increase in demand for ubiquitous and high data-rate wireless communications. To cope with this exponentially growing rate, it is expected that cellular wireless systems would need $100\times$ increase in Spectral Efficiency~(SE) and $1000\times$ improvement in Energy Efficiency~(EE) by 2020, which calls for a technological revolution. While the current cellular network architecture was not originally designed for such capabilities, Cloud Radio Access Network~(C-RAN)~\cite{whitepaper13} has been introduced recently as a revolutionary redesign of the cellular architecture to address the huge increase in data traffic and to reduce the capital expenditure~(CAPEX) and operating expenditure~(OPEX)~\cite{wu2015cloud}. 

Cloud Radio Access Network~(C-RAN)~\cite{whitepaper13} has been introduced as a revolutionary redesign of the cellular architecture to address the increase in data traffic and to reduce the capital expenditure~(CAPEX) and operating expenditure~(OPEX)~\cite{wu2015cloud}. The idea of C-RAN is to decouple the computational functionalities from the distributed BS (a.k.a. eNodeB in LTE) and to consolidate them in a centralized processing center. A typical C-RAN is composed of: (i)~light-weight, distributed Radio Remote Heads~(RRHs) plus antennae, which are located at the remote site and are controlled by a centralized virtual base station pool, (ii)~the Base Band Unit~(BBU) composed of high-speed programmable processors and real-time virtualization technology to carry out the digital processing tasks, and (iii)~low-latency high-bandwidth optical fibers, which connect the RRHs to the BBU pool. In a centralized BBU pool, since information about the network resides in a common place, the BBU can exchange control data at $\mathrm{Gbps}$ rate. By exploiting the global view of the network condition and traffic demand available at the BBU, dynamic provisioning and allocation of spectrum, computing, and radio resources can improve network performance~\cite{tran2015mass, pompili2016elastic, tran2016mass, tran2017twireless, luong2017fast, hosseini2016cloud}. \emph{Interestingly, C-RAN paves the way for bridging the gap between two so-far disconnected worlds: cellular communications and cloud computing.}

% This centralized characteristic---along with virtualization technology and low-cost relay-like RRHs---provides a higher degree of freedom to make optimized decisions and has made C-RAN a promising technology candidate to be incorporated into the Fifth Generation~(5G) wireless network, especially for urban/high-density areas. For instance, 

% based on the global view of the network condition and traffic demand available at the BBU pool, dynamic provisioning and allocation of spectrum, computing, and radio resources can improve network performance~\cite{tran2015mass, pompili2016elastic, tran2016mass, tran2017twireless, luong2017fast}. \emph{In some respect, C-RAN paves the way for bridging the gap between two so-far disconnected worlds: cellular communications and cloud computing.}

In a BBU pool, most of the communication functionalities are implemented in part or fully in a virtualized environment hosted over general-purpose computing servers and can be realized using Virtual Machines~(VMs).
% that are housed in one or more racks in a nearby cloud datacenter. 
% It is therefore crucial to design and provision the virtualized environment properly in order to make it flexible and energy efficient while also capable of handling intensive computations. 
% Such a virtualized environment can be realized via the use of Virtual Machines~(VMs). 
The flexible reconfigurability of the virtualized BBU allows for it to be dynamically resized `on the fly' in order to meet the fluctuations in capacity demands. This \emph{elasticity} will enable significant improvement in user Quality of Service~(QoS) and efficiency in energy and computing resource utilization in C-RANs. However, determining the computational resources of a virtualized BBU~(VM) that is capable of providing adequate processing capabilities with respect to the traffic load presents non-trivial challenges. 

\textbf{Our vision:}
We seek to characterize the computational requirements of virtualized BBUs over a real-world implementation of a small-scale C-RAN system. Software implementations coupled with real-hardware experiments is essential to understand the runtime complexity as well as performance limits of the BBU in terms of processing throughput and latency and how they translate into mobile-user QoS metrics. The realization of the C-RAN emulation testbed on virtualized general-purpose computing servers will allow for \emph{profiling} of the computational complexity of the different communication functionalities implemented in software. In particular, such profiling results will provide a ``mapping'' from the number and combination of different types of user traffic to VM computational capacity. \emph{Hence, we aim at establishing empirical models for the estimation of processing time and CPU utilization with respect to different radio-resource configurations and traffic load}. Our model will provide researchers and practitioners with real-world insights and the necessary tools for designing advanced and efficient resource-provisioning and allocation strategies in C-RAN systems.

\textbf{Related works:}
There has been a considerable number of works addressing the benefits of C-RAN from the cooperative communications perspectives. For instance, the work in~\cite{luong2016joint} consider the power minimization problem by jointly optimizing the set of active RRHs and precoding or beamforming design. 
% The considered power models consist of the RRH transmission power~\cite{vu2015tvt}, and additionally the user transmission power in~\cite{Luo2015downlink}, transport network power in~\cite{shi2014group}, and power consumption at the BBU pool in~\cite{TanTayQueJ15}. 
In addition, the optimal tradeoff between transmission power and delay performance is investigated in~\cite{tran2016secon} via a cross-layer based approach, taking into account the imperfect channel information. Furthermore, the works in~\cite{vuadaptive, nguyen2017energy} address the front-haul uplink compression problem in C-RAN. While showing promising performance gains brought by the centralized cooperation and network optimization C-RAN, these works often overlook the system issues and mostly rely on simplified assumptions when modeling the computational resources of the BBU.   
From the system perspectives, several LTE RAN prototypes have been implemented over General-Purpose Platforms~(GPPs) such as the Intel solutions based on hybrid GPP-accelerator~\cite{SchoolerIntel}, Amarisoft solution~\cite{amarisoft}, and OpenAirInterface platform~\cite{oai}. Studies on these systems have demonstrated the preliminary potential benefits of C-RAN in improving statistical multiplexing gains, energy efficiency, and computing resource utilization. Field-trial results in~\cite{whitepaper13, ChihLin_RecentProg} show the feasibility of deploying C-RAN front-haul using CPRI compression, single fiber bidirection, and wavelength-division multiplexing. The authors in~\cite{mao2015minimizing} focus on the issue of minimizing computational and networking latencies by VMs or containers. Kong et al.~\cite{kong2013ebase} present the architecture and implementation of a BBU cluster testbed to improve energy efficiency in C-RAN. Wu~\cite{wu2014pran} shows a high-level architecture for programmable RAN~(PRAN) that centralizes base stations' L1/L2 processing of BBU pool onto cluster of commodity servers. This approach shows the feasibility of fast data path control and efficiency of resource pooling. In summary, these works focus on the overall system architecture, feasibility of virtual software BS stacks, performance requirements, and analysis of optical links between the RRHs and the BBU cloud. However, most of these systems are either proprietary or ad-hoc based, and do not provide a generalized characterization that facilitates research on new algorithms. 

\textbf{Our contributions:}
Given the importance of designing effective resource-management solutions in C-RAN and the lack of experimental studies into the computational performance and requirements of the BBU pool, we make the following contributions in this paper.
\begin{itemize}[leftmargin=*]
\item We present the design and implementation of a programmable C-RAN testbed comprising of a virtualized BBU connected to multiple eNodeBs~(eNBs). In particular, the BBU is implemented using an open-source software platform OpenAirInterface~(OAI) that allows for simulation and emulation of the LTE protocol stack. The eNBs are realized using programmable USRP boards. 

\begin{figure}[!ht]
 \centering
 \includegraphics[width=0.44\textwidth]{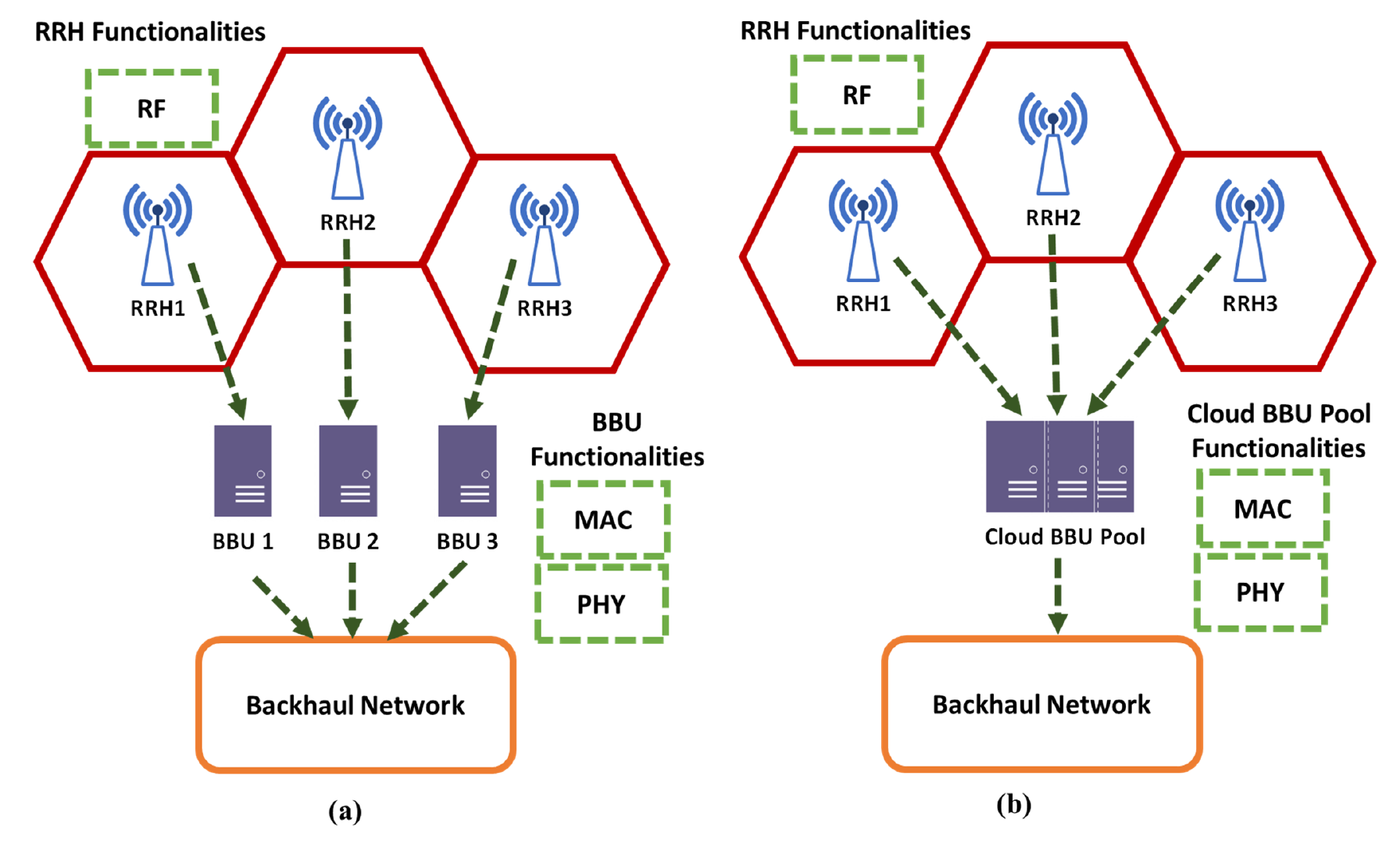}
\caption{(a) Each BBU is assigned to one RRH, (b) Consolidated BBU.}\label{C-RAN}
\end{figure}

\item We perform extensive experiments with transmissions between the eNB and the UE under various configurations in order to profile the runtime complexity and performance limits of the BBU in terms of processing, throughput, and latency. It is shown that the processing time and CPU utilization of the BBU increase with the Modulation and Coding Scheme~(MCS) index and with the number of allocated Physical Resource Blocks~(PRBs).

\item Using empirical data, we model the BBU processing time as a function of the CPU frequency, MCS, and PRBs, and show that the CPU utilization percentage of the BBU can be well approximated as a linear increasing function of the maximum downlink data rate. These approximated models provide real-world insights and key inputs to formulate, design, and evaluate optimized resource-management problems in C-RAN. 
\end{itemize}

\textbf{Paper organization:}
%The remainder of this paper is organized as follows. 
In Sect.~\ref{sec:mode2}, we describe the C-RAN system architecture; in Sect.~\ref{sec:mod3}, we discuss the design and implementation of our C-RAN testbed; in Sect.~\ref{sec:results}, we present the experimental results and provide our empirical models;  finally, we conclude the paper in Sect.~\ref{sec:conclusion}.

%---------------------------------------------------   
\section{System Overview}\label{sec:mode2}
We describe here the C-RAN system architecture and the OAI software platform that is capable of realizing a virtualized C-RAN system. 

% We then discuss the critical issues related to the implementation of virtualized C-RAN. 

\subsection{C-RAN Architecture}
The general architecture of C-RAN mainly consists of two parts: the distributed RRHs plus antennae deployed at the remote site and the centralized BBU pool hosted in a cloud datacenter. The BBU pool consists of multiple BBUs, each hosted on a VM and connected to the corresponding RRH via high-bandwidth low-latency media (e.g., use of optic fibers allows for maximum distance separation of $40~\rm{Km}$ between the RRH and its BBU~\cite{whitepaper13}). Packet-level processing, Medium Access Control~(MAC), physical-layer~(PHY) baseband processing, and Radio Frequency~(RF) functionalities may be split between the BBU and the RRHs depending on the specific C-RAN implementation. In this paper, we consider the full centralization of C-RAN in order to exploit fully the potential of this paradigm where only RF functionalities are deployed at the RRHs. Based on the network performance and system implementation complexity, each BBU can be assigned to one RRH, as shown in Fig.~\ref{C-RAN}(a) or the BBUs can be consolidated into one entity, called BBU pool as depicted in Fig.~\ref{C-RAN}(b).

\begin{figure}[!ht]
\centering
 \includegraphics[width=0.44\textwidth]{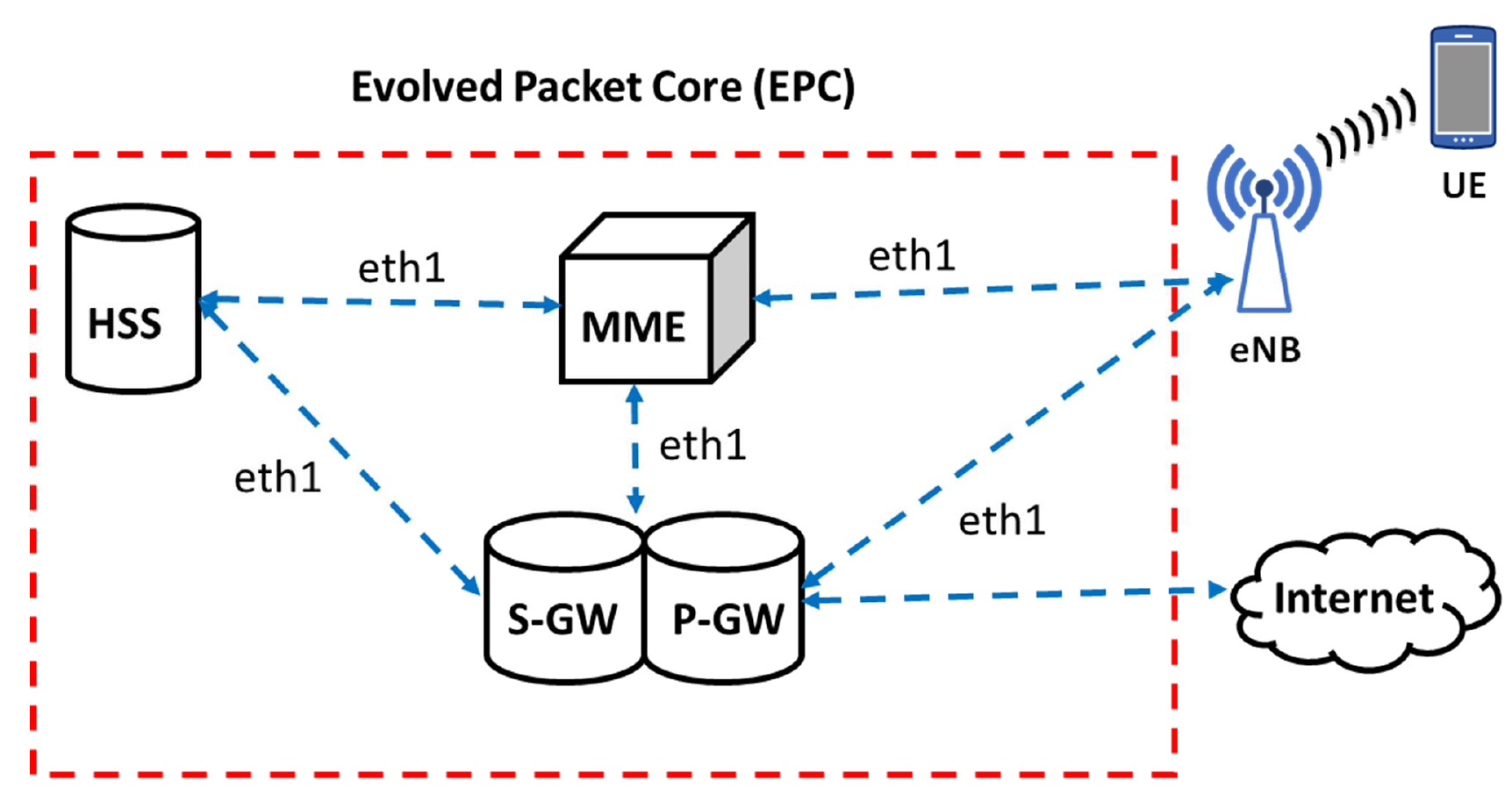}
\caption{Evolved Packet Core~(EPC) network topology diagram.}\label{EPC}
\end{figure}

\subsection{Emulation Platform}
We choose an open-source software implementation of LTE standard called OpenAirInterface~(OAI)~\cite{oai} developed by EUROCOM to realize the virtualized C-RAN system. OAI can be used to build and customize mobile network operators consisting of eNBs and Commercial off-the-shelf~(COST) UEs as well as software-defined UEs.
% This is a wireless communications platform developed by EUROCOM to provide a complete flexible cellular ecosystem towards an open-source 5G implementation. OAI can be used to build and customize mobile network operators consisting of eNBs and Commercial off-the-shelf~(COST) UEs as well as software-defined UEs. In addition, OAI offers facilities to configure and monitor the RAN in real time via a software radio front-end connected to a host computer for processing. This approach is similar to other Software-Defined Radio~(SDR) prototyping platforms in the wireless networking research community such as OpenBTS~\cite{burgess2008openbts}. 
The structure of OAI mainly consists of two components: one part, called $\rm{Openairintereface5g}$, is used for building and running eNB units; the other part, called $\rm{Openair}$-$\rm{cn}$, is responsible for building and running the Evolved Packet Core~(EPC) networks, as shown in Fig.~\ref{EPC}. The $\rm{Openair}$-$\rm{cn}$ component provides a programmable environment to implement and manage the following network elements: Mobility Management Entity~(MME), Home Subscriber Server~(HSS), Serving Gateway~(S-GW), and PDN Gateway~(P-GW).

\begin{figure*}[t]
 \centering
 \begin{tabular}{ccc}
\hspace*{-.3cm}\includegraphics[width=0.25\textwidth]{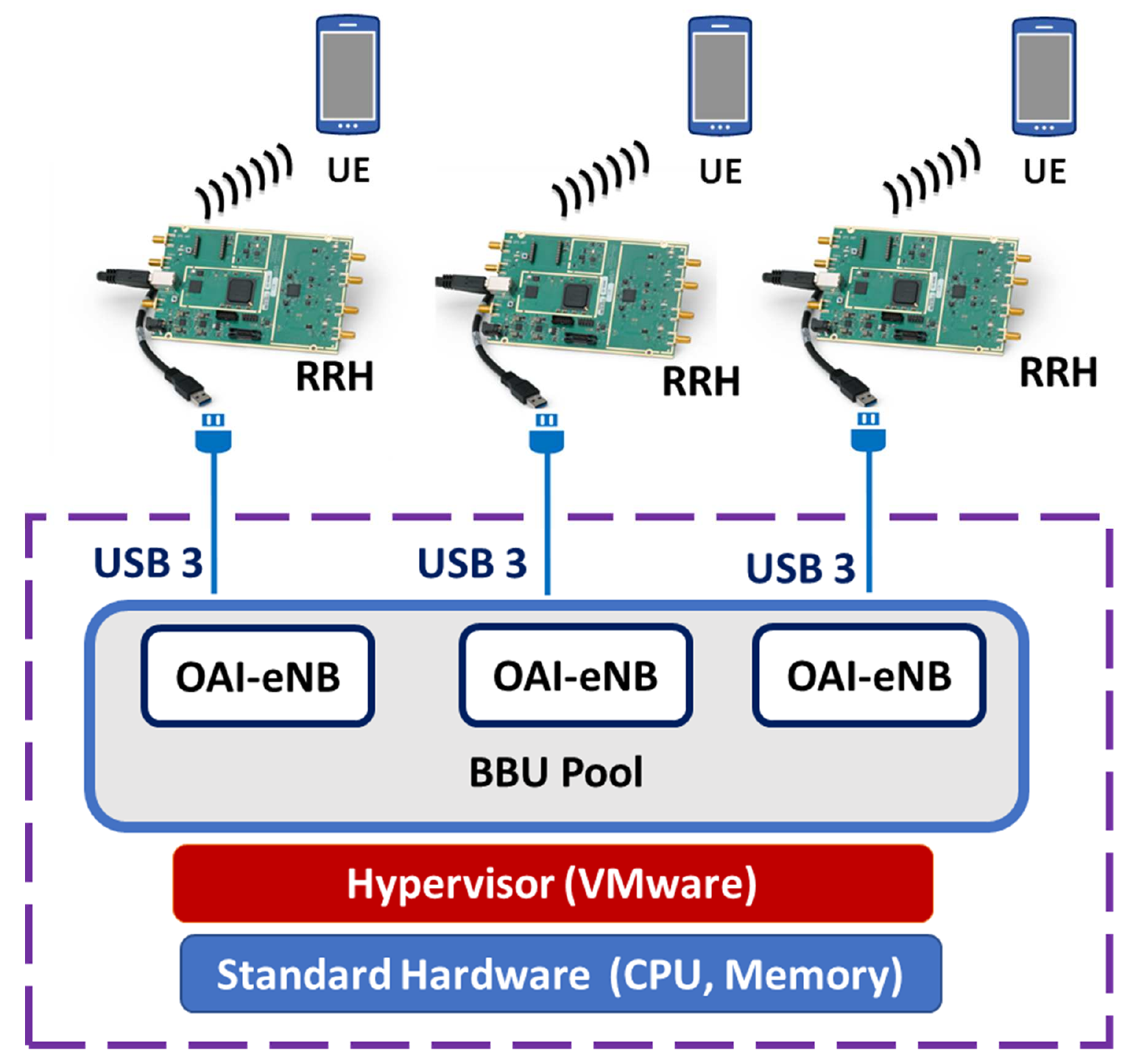} &
\hspace*{.5cm}\includegraphics[width=0.3\textwidth]{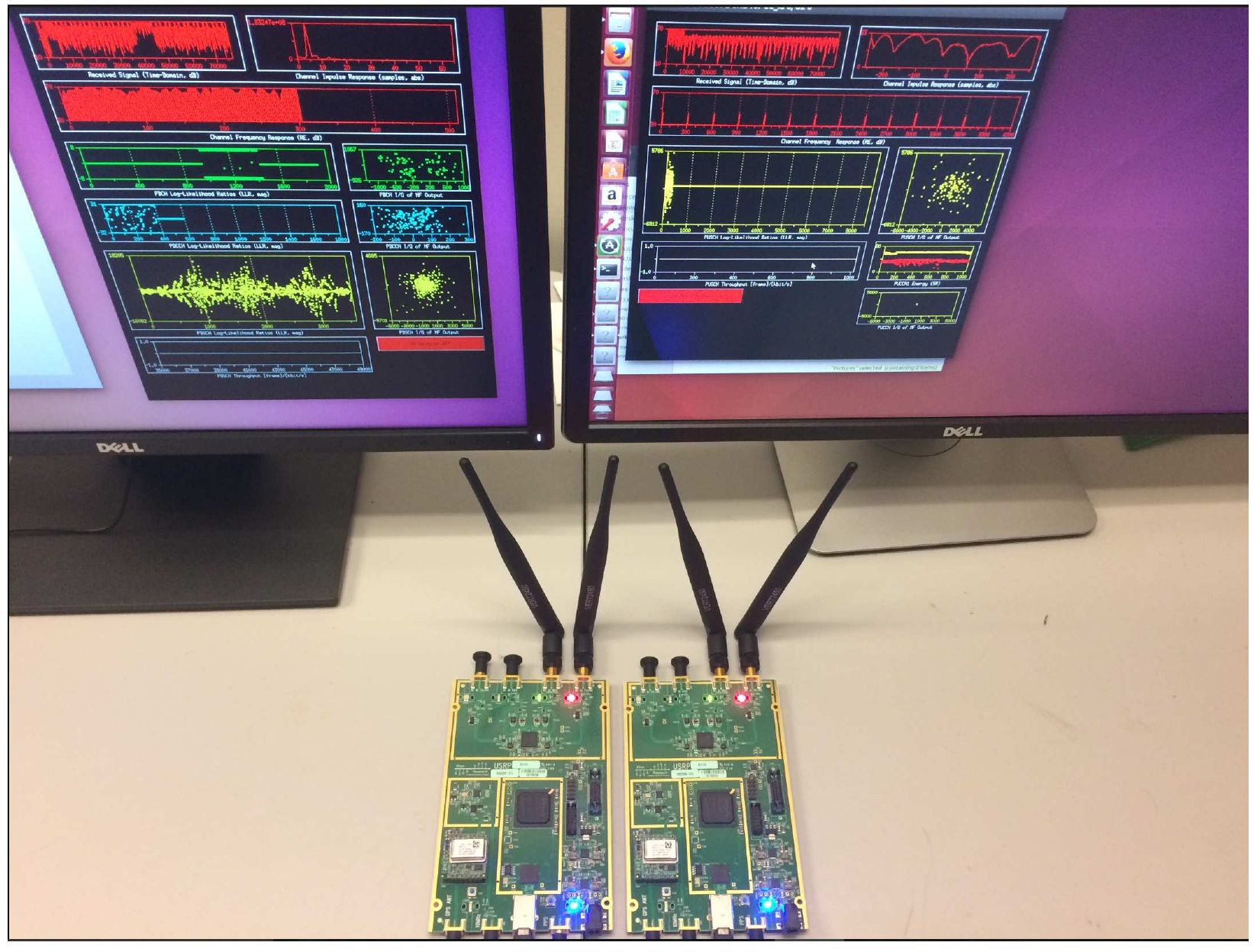} &
\hspace*{.5cm}\includegraphics[width=0.28\textwidth]{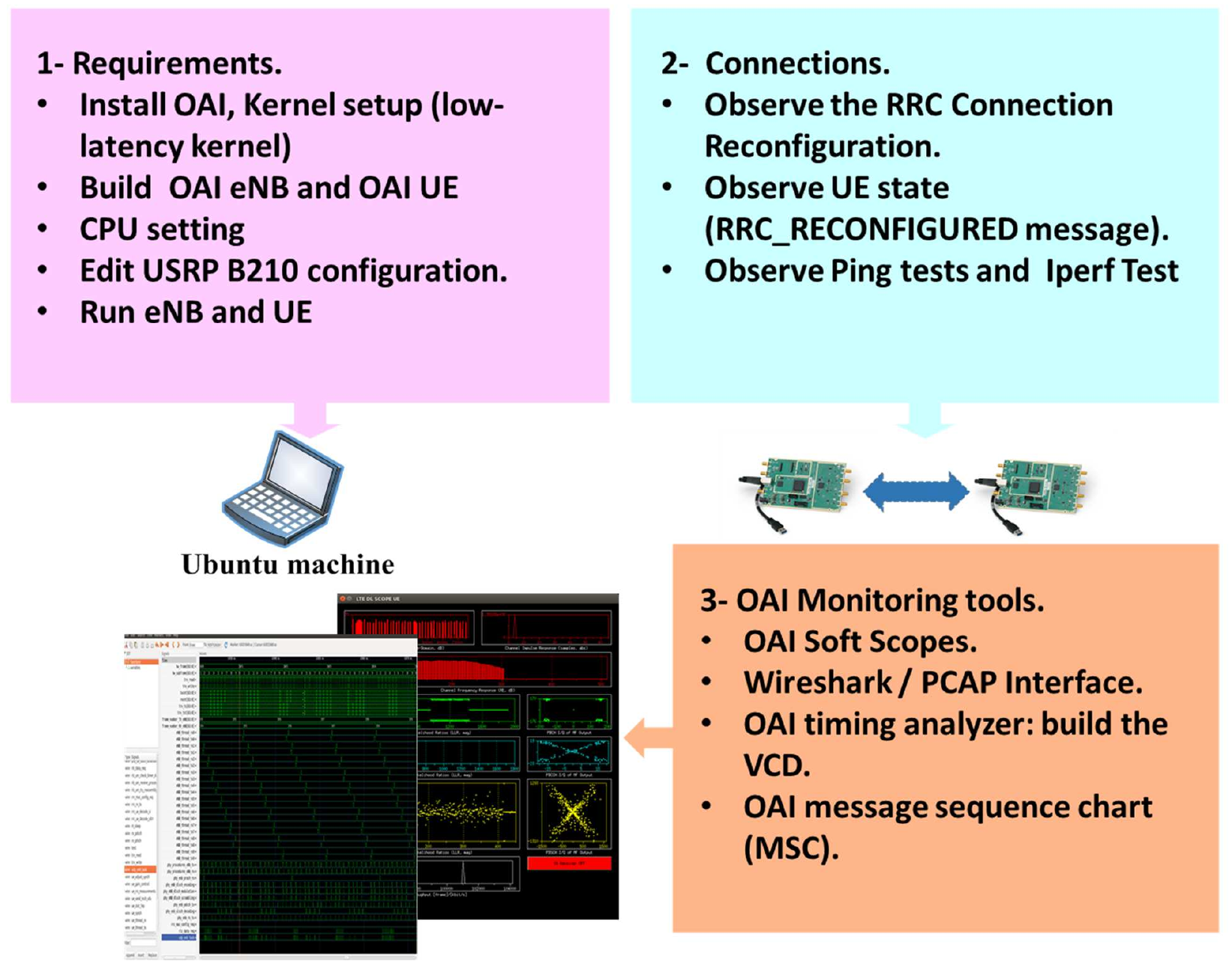} \\
 \small(a) & \small(b) & \small(c)
\end{tabular}
\caption{(a) Logical illustration of C-RAN testbed architecture; (b) C-RAN testbed implementation utilizing OAI; and (c) OAI processing flow. 
}\label{fig:testbed3}
\end{figure*}
%---------------------------------------------------   
\section{C-RAN Experimental Testbed}\label{sec:mod3}
We detail now our C-RAN testbed using OAI, including the testbed architecture, configuration, and experiment methods. 

\subsection{Testbed Architecture}
Figure~\ref{fig:testbed3}(a) illustrates the architecture of our testbed. The RRH front-ends of the C-RAN testbed are implemented using SDR USRP B210s, each supporting $2\times2$ MIMO with sample rate up to $62~\rm{MS/s}$. In addition, each radio head is equipped with a GPSDO module for precise synchronization. Each instance of the virtual BBU is implemented using the OAI LTE stack, which is hosted in a VMware VM. All the RRHs are connected to the BBU pool (the physical servers hosting the VMs) via USB~$3$ connections. 

The Ubuntu $\rm{14.04}$ LTS with kernel $\rm{3.19.0}$-$\rm{91}$-$\rm{lowlatency}$ is used for both host and guest operating systems. In order to achieve a high performance for our testbed, all power-management features in the BIOS, C-states, and CPU frequency scaling have been turned off. The CPU should support the $\rm{sse3}$ and $\rm{sse4.1}$ features. These flags must be exposed from the host to the guest, and can be checked by using the command $\rm{cat/proc/cpuinfo|grep}\,\,flags|uniq$. For the physical sever hosting the BBU, we use a Dell Precision T5810 workstation with Intel Xeon CPU E5-1650, 12-core at $3.5~\rm{GHz}$, and $32~\rm{GB}$ RAM. There are several configurations that depend on the guest OS's specific setup that should be calibrated in order to boost the performance of the testbed. Most importantly, the maximum transmit power at the eNB and the UE can be calibrated as follows.

\begin{itemize}[leftmargin=*]
\item \textbf{eNB:} The maximum transmit power at the eNB is signaled to the UE so that it can do its power control. The parameter is PDSCH Energy Per Resource Element~(EPRE)~$\rm[dBm]$ and it is part of the configuration file, $\rm{pdsch\_referenceSignalPower}$. It should be measured using a vector signal analyzer with LTE option for the utilized frequency and then put in the configuration file.

\item \textbf{UE:} At the UE, the maximum transmit power~$\rm[dBm]$ is measured over the whole (usable) bandwidth. If the same hardware is used at the UE and at the eNB, the power is ${\rm{max\_ue\_power}} = {\rm{PDSCH\_EPRE}} + 10\log_{10}\left( {12{\rm{N\_PRB}}} \right)$.
\end{itemize}

%%%%%%%%%%%%%%%%%%%%%%%%%%%%%%%

% \begin{figure}[!ht]
%  \centering
%  \includegraphics[width=0.44\textwidth]{fig/testbed.eps}
% \caption{C-RAN experimental testbed utilizing OAI.}\label{tesbed}
% \end{figure}

\subsection{Monitoring the OAI eNB and the UE}
As illustrated in Fig.~\ref{fig:testbed3}(b), our C-RAN experimental testbed consists of one unit of UE and one unit of eNB, both implemented using the USRP B210 boards and running on OAI. The OAI software instances of the eNB and UE run in separate Linux-based Intel x86-64 machines comprising of 4 cores for UE and 12 cores for eNB, respectively, with Intel i7 processor core at $3.6~\rm{GHz}$. 

OAI comes with useful monitoring tools such as network protocol analyzers, loggers, performance profilers, timing analyzers, and command line interfaces for performing the intended measurements and monitoring of the network. Specifically, the supported monitoring tools include:

\begin{itemize}[leftmargin=*]
\item OAI Soft Scope, which monitors received-transmitted waveforms and also tracks the channel impulse response.

\item WireShark Interface and ITTI Analyzer, which can be used to analyze the exchanges between eNB and UE protocols.
 
 \item OpenAirInterface performance profiler, which is used for processing-time measurements.
 
%  \item OAI tools such as network protocol analyzers, loggers, performance profilers, timing analyzers and command line interfaces which use for performing the intended measurements and monitoring of the network. 
\end{itemize}

\begin{table}[htb!] 
\renewcommand{\arraystretch}{1.5}
\caption{Testbed Configuration Parameters}
\centering \label{tab:para}
\begin{tabular}{|c|c|c|} 
\hline
\textbf{Parameters} &  \textbf{eNB} & \textbf{UE} \\
\hline
Duplexing  mode & FDD & FDD \\
\hline
Frequency  &  $  2.66~\rm{GHz}$ & $ 2.66~\rm{GHz}$ \\
\hline
Transmitted power  &  $[150 \div 170]~\rm{dBm}$ & $[150 \div 170]~\rm{dBm}$ \\
\hline
MCS & $[0 \div 27]$ & $[0 \div 27]$\\
\hline
Mobility & Static &	Static \\
\hline 
PRB &	$25,50,100$	& $25,50,100$ \\
\hline
Radiation pattern &	Isotropic &	Isotropic \\
\hline
\end{tabular}
\end{table}

Figure~\ref{fig:testbed3}(c) illustrates the OAI processing flow for building, running, and monitoring stages. In addition, we summarize the testbed configuration parameters in Table~\ref{tab:para}. In particular, the  eNB is configured in band 7~(FDD)
% using a downlink~(DL) carrier frequency of $2.66~\mathrm{GHz}$. The 
and the transmission bandwidth can be set to $5$, $10$, and $20~\rm{MHz}$, corresponding to $25$, $50$, and $100~\rm{PRBs}$, respectively. In order to determine the successful connection between eNB and UE, the RRC states should be observed in OAI console. Specifically, when the UE is successfully paired to the eNB, the RRC connection setup message should be seen.

\subsection{Interference-free Testbed Environment}
\begin{figure}[!ht]
 \centering
 \includegraphics[width=0.4\textwidth]{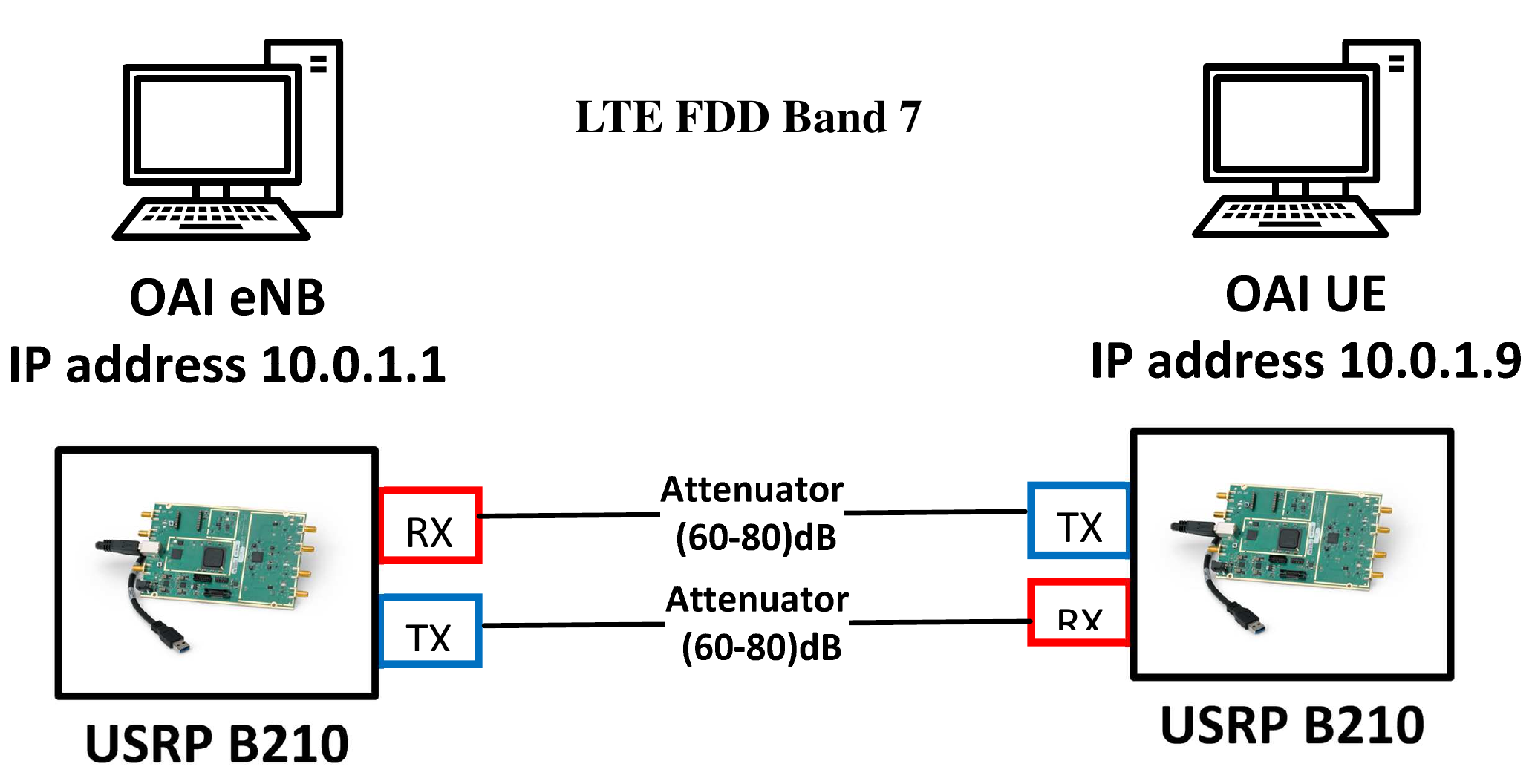}
\caption{Configuration of the eNB-UE connection.}\label{Atest}
\end{figure}

We set up the experiment environment to emulate a ``quiet'' transmission between the eNB and UE in which there is no interference from other devices (so to have better control of the environment). To accomplish this, we use two configurable attenuators, model name Trilithic Asia 35110D-SMA-R, which connect the Tx and Rx ports of the eNB to the Rx and Tx ports of the UE, respectively, as shown in Fig.~\ref{Atest}.  In order to establish a stable connection, the transmitter and received gains in the downlink have been set to $90$ and $125~\rm{dB}$, respectively.

\begin{figure}[!ht]
 \centering
 \includegraphics[width=0.4\textwidth]{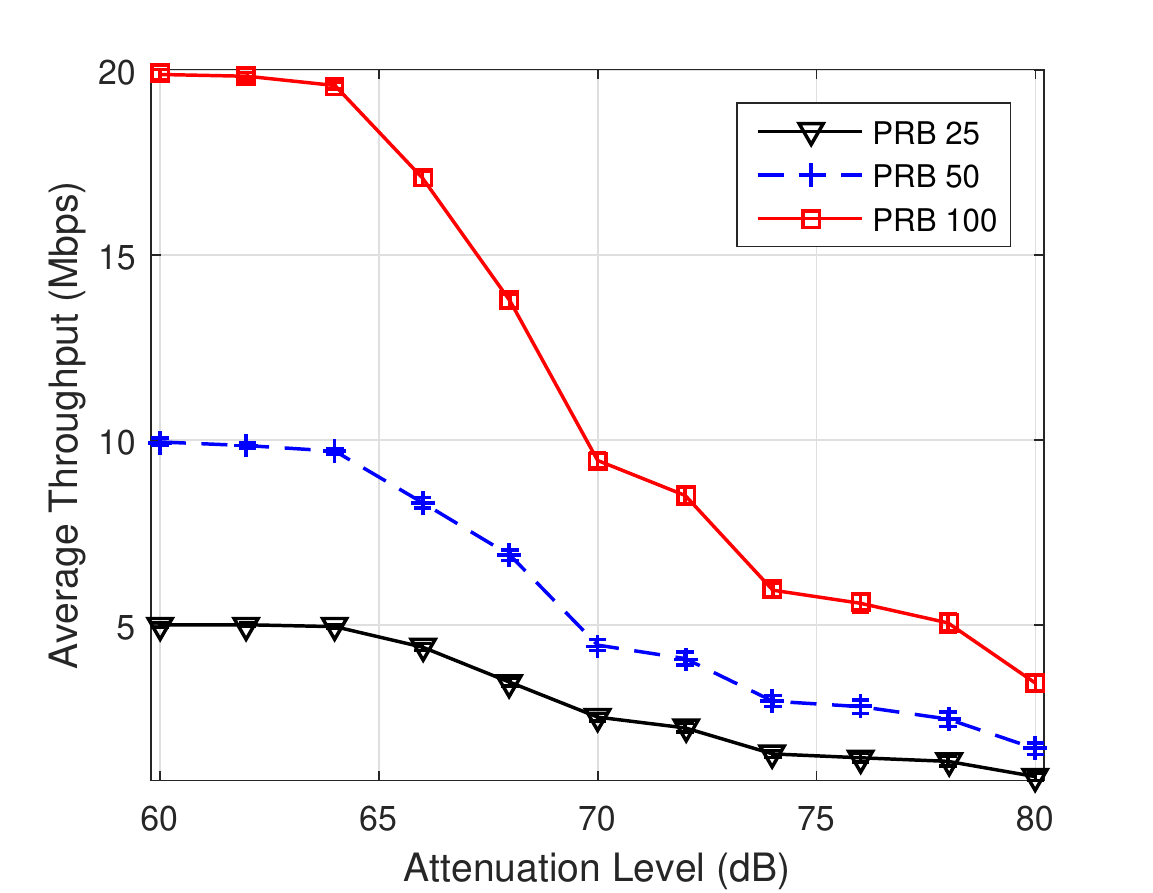}
\caption{Downlink throughput performance at different attenuation levels.}\label{fig:attenuation}
\end{figure}

We use $\rm{iperf}$ to generate $500$ packets to send from the eNB to the UE. Figure~\ref{fig:attenuation} illustrates the throughput performance versus the attenuation level between the eNB and UE. 
% The attenuation was varied between $60$ and $80~\rm{dB}$. 
It can be seen that the achievable throughput significantly decreases with attenuation level. Specifically, when the attenuation level is $60~\rm{dB}$ the achievable throughputs are around $5$, $10$, and $20~\rm{Mbps}$ when using $25$, $50$, and $100$ PRBs, respectively. On the other hand, at an attenuation of $80~\rm{dB}$, the throughputs are much lower, i.e., $0.98$, $1.64$, and $3.40~\rm{Mbps}$, respectively. We observe that the eNB-UE connection will drop when the attenuation level goes higher than $80~\rm{dB}$. 
% The results in Fig.~\ref{fig:attenuation} show that when the attenuation level is $60~\rm{dB}$ the achievable throughputs are around $5$, $10$, and $20~\rm{Mbps}$ when using $25$, $50$, and $100$ PRBs, respectively. On the other hand, at an attenuation of $80~\rm{dB}$, the throughputs are much lower, i.e., $0.98$, $1.64$, and $3.40~\rm{Mbps}$, respectively.      

%The extinction time for functions~(PHY, MAC, RLC, PDCP, and GTP) has been tracked by using the GTKWave, a software used to read VCD file generated by the OAI software. The GTKWave can be installed in Linux OS and run using the command gtkwave. The PDCP is protocol sitting between the RLC and RRC layers in the LTE air interface to connect an UE with an eNB. The CPU spends most of its time, as illustrated in Fig.~\ref{CPU}, on PHY channel processing for Band 5 and Band 7.

\section{Experimental Results and Empirical Models}\label{sec:results}
We present the performance of the virtualized BBU, i.e., the OAI eNB, in terms of packet delay, CPU processing time, and utilization under various PRB and MCS configurations. 

\subsection{Delay Performance}
\begin{figure}[!ht]
 \centering
 \includegraphics[width=0.4\textwidth]{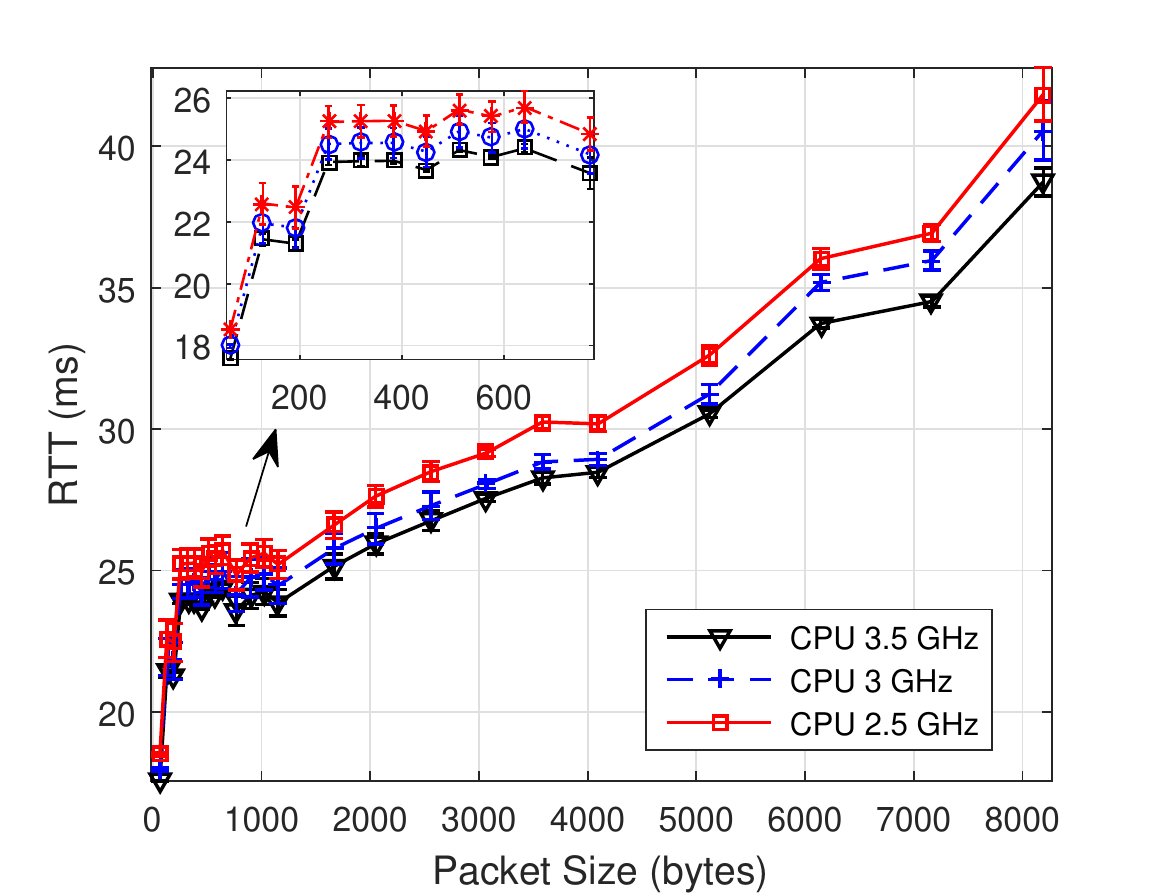}
\caption{RTT measurement for different packet sizes.}\label{Paket-size}
\end{figure}

To test the delay in the C-RAN testbed, we focus on measuring the RTT when sending packets between the eNB and the UE. The VM hosting the BBU is configured with $4$ virtual cores and $8~\rm{GB}$ RAM in a VMware hypervisor, running on a physical machine with $12$ cores, $3.5~\rm{GHz}$ CPU, and $16~\rm{GB}$ RAM. The OAI UE runs on a low-latency Ubuntu physical machine with $3.0~\rm{GHz}$ CPU and $8~\rm{GB}$ RAM. Figure~\ref{Paket-size} illustrates the relationship between RTT and packet size when the BBU is set at different CPU frequencies. For each experiment, we sent $500$ Internet Control Message Protocol~(ICMP) echo request packets from the eNB to the UE. It can be seen that the RTT exponentially increases as the packet size increases. Moreover, we have also noted that the RTT is greater when OAI eNB runs on a VM than on a physical machine, which may be due to the overhead incurred when running the VM. In addition, there is a correlation between the CPU frequency and the OAI software performance. We have recorded that the minimum CPU threshold frequency to run OAI in our scenario is $2.5~\rm{GHz}$. Below the threshold value, we observed that the synchronization between eNB and UE is occasionally missed. By controlling the CPU frequency using the $\rm{Cpupower}$ tool, we have noticed that the RTT can be improved by increasing the CPU frequency steps. 

%\begin{figure}[!ht]
% \centering
% \includegraphics[width=0.45\textwidth]{fig/Attenuation.eps}
%\caption{Downlink throughput performance at different attenuation levels.}\label{Attenuation}
%\end{figure}

% \begin{figure}[!ht]
% \centering
% \includegraphics[width=0.45\textwidth]{fig/Paket-size.eps}
%\caption{RTT measurement for different packet sizes.}\label{Paket-size}
%\end{figure}

\subsection{Processing Time of LTE Subframes}
In this section, we study the BBU processing time of each LTE subframe with respect to different CPU frequency configurations in the VMware environment. The execution time of each signal processing module in the downlink is measured using \emph{timestamps} at the beginning and at the end of each subframe. OAI uses the RDTSC instruction implemented on all x86 and x64 processors as of the Pentium processors to achieve precise timestamps~\cite{alyafawi2015critical}. The $\rm{cpupower}$ tool in Linux is used to control the available CPU frequencies. To avoid significant delay and to not miss the synchronization between eNB and UE hardware, we recommend to run the experiment within a $2.8 \div 3.5~\mathrm{GHz}$ CPU frequency range. 

\begin{figure}[!ht]
 \centering
 \includegraphics[width=0.4\textwidth]{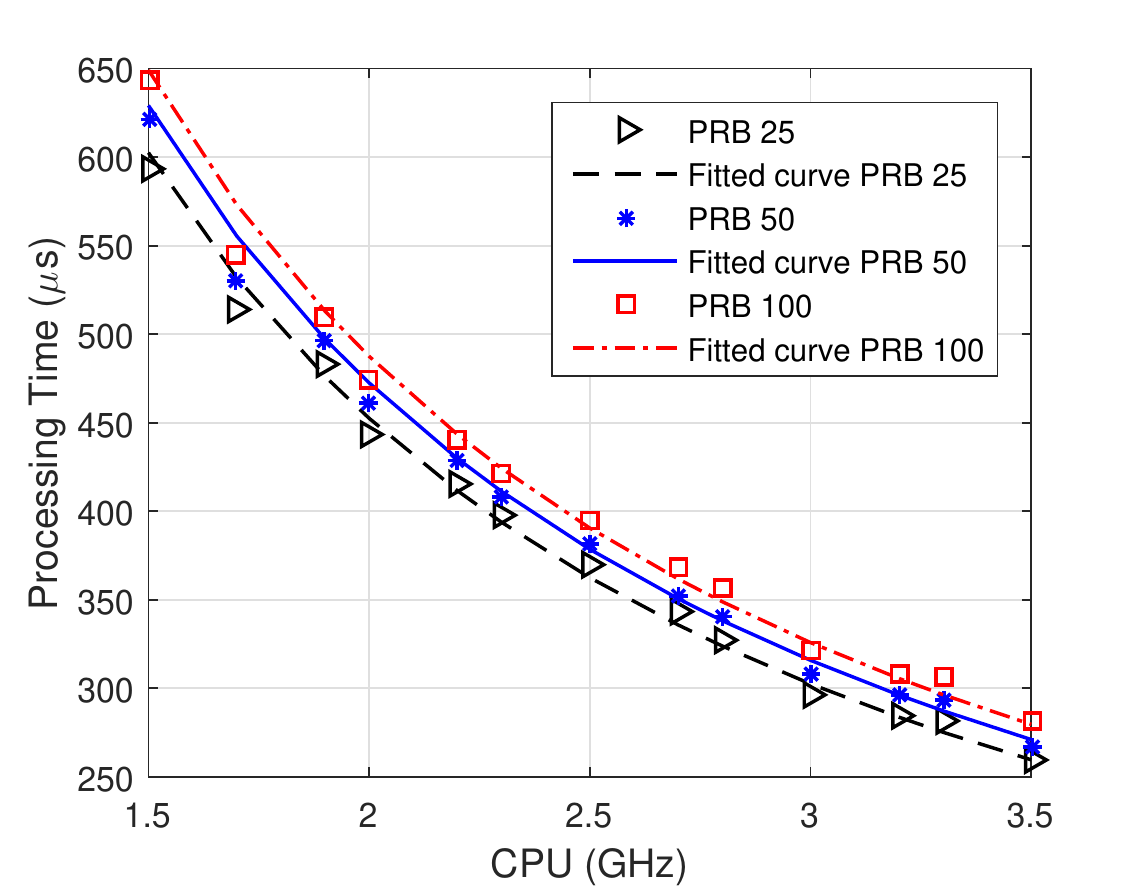}
\caption{Processing time of LTE subframes against CPU frequency with $\rm{MCS} =27$ and various $\rm{PRB}$ allocations.}\label{BBU}
\end{figure}

In Fig.~\ref{BBU}, we depict the processing time of the eNB given different CPU-frequency steps, in which the MCS index is set to $27$ for both UL and DL. It can be seen that the processing time dramatically decreases when the CPU frequency increases. To model the subframe processing time against the CPU frequency and radio-resource configuration, we repeat the experiment in Fig.~\ref{BBU} with different MCS indexes. The subframe processing time ${T_{{\rm{sub}}}}\left[ {\mu s} \right]$ can be well approximated as a function of CPU frequency $f~\rm[Hz]$, MCS, and PRB as,
%\begin{equation}
%{T_{{\rm{sub}}}}\left[ {\mu s} \right] = \alpha /f + {\delta _{{\rm{PRB}}}} + {\psi _{{\rm{MCS}}}},
%\end{equation}
\begin{equation}
{T_{{\rm{sub}}}}\left[ {\mu s} \right] = \frac{{{\alpha _{{\rm{PRB}}}}}}{f} + {\beta _{{\rm{MCS}}}} + 2.508,
\end{equation}
where ${\alpha _{{\rm{PRB}}}}$ and ${\beta _{{\rm{MCS}}}}$ are two parameters that increase with PRB and MCS values as reported in Table~\ref{tab:PRB}.
%Tables~\ref{tab:aPRB} and \ref{tab:bMCS}, respectively.

% \begin{table}[h!]
% \renewcommand{\arraystretch}{1.5}
% \caption{Values of ${\alpha _{{\rm{PRB}}}}$}\label{tab:aPRB}
% \centering
% \begin{tabular}{|c|c|c|c|}
% \hline
% PRB & 25 & 50 & 100 \\
% \hline
% ${\alpha _{{\rm{PRB}}}}[\mu s]$ & 900 & 940 & 970 \\
% \hline
% \end{tabular}
% \end{table}

% \begin{table}[h!]
% \renewcommand{\arraystretch}{1.5}
% \caption{Values of ${\beta _{{\rm{MCS}}}}$}\label{tab:bMCS}
% \centering
% \begin{tabular}{|c|c|c|c|c|c|c|c|}
% \hline
% MCS & 0 & 9 & 10 & 16 & 17 & 24 & 27\\
% \hline
% ${{\beta}_{\rm{MCS}}}[\mu s]$ & 0 & 9.7 & 11.8 & 37.5 & 39.7 & 64.8 & 75\\
% \hline
% \end{tabular}
% \end{table}

\begin{table}[h!]
\caption{Values of parameters ${\alpha _{{\rm{PRB}}}}$ and ${\beta _{{\rm{MCS}}}}$.}\label{tab:PRB}
\centering
 \includegraphics[width=0.4\textwidth]{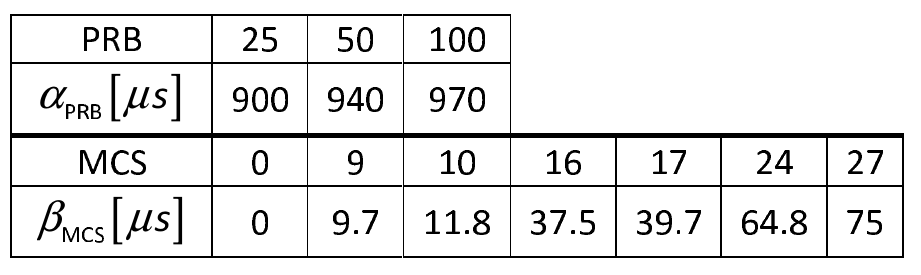}
\end{table}

\subsection{CPU Utilization}
\begin{figure}[!ht]
 \centering
 \includegraphics[width=0.4\textwidth]{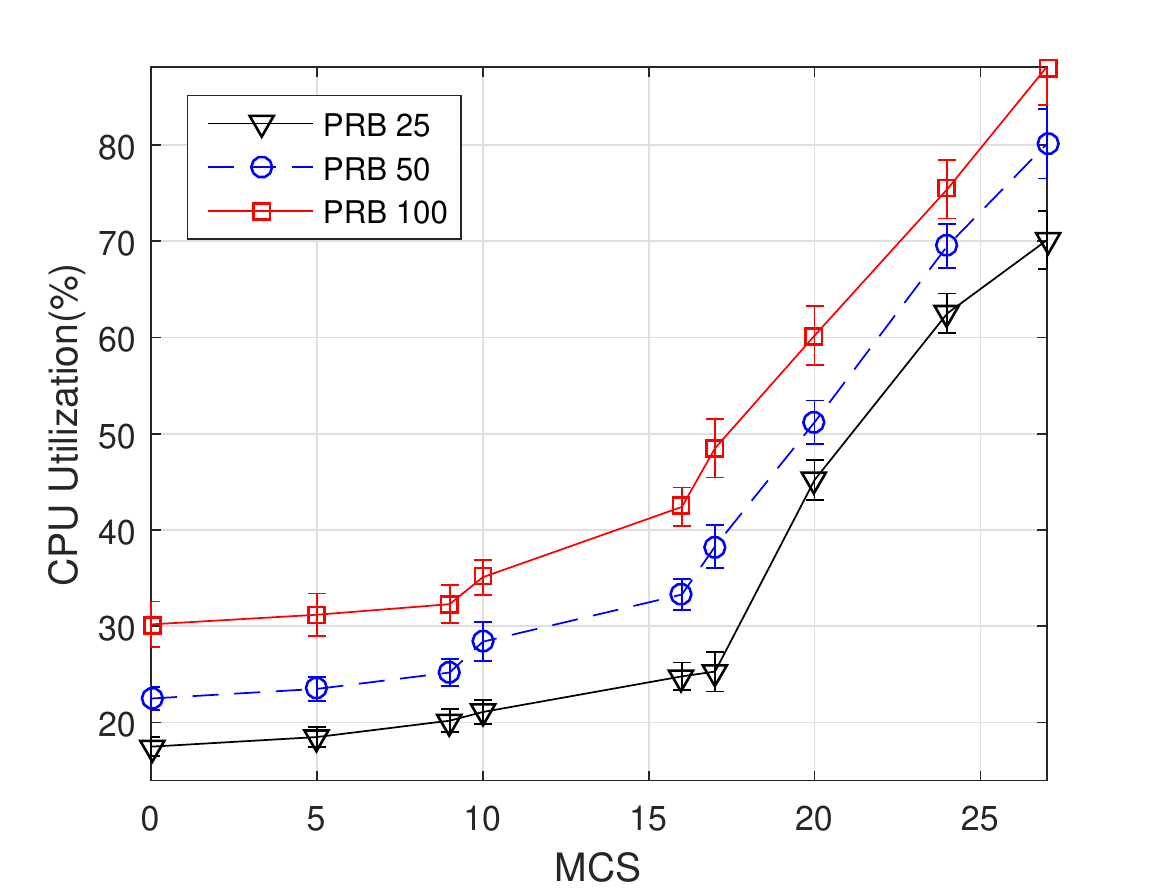}
\caption{CPU utilization of the BBU at different values of MCS and PRB.}\label{CPU_per}
\end{figure}

In C-RAN, it is of critical important to understand the CPU utilization of the BBU in order to design efficient resource provisioning and allocation schemes. In the previous subsections, we have seen the relationship between MCS and CPU usage for different values of PRBs. In this experiment, the CPU utilization percentage is calculated using the $\rm{top}$ command in Linux, which is widely used to display processor activities as well as various tasks managed by the $\rm{kernel}$ in real time. We repeatedly send UDP traffic from the eNB to the UE with various MCS and PRB settings. The CPU utilization percentage has been recorded as in Fig.~\ref{CPU_per}. By setting the CPU frequency of the OAI eNB to $3.5~\rm{GHz}$, we have seen that the highest CPU consumption occurred at MCS $27$, corresponding to $72\%$, $80\%$, and $88\%$ when PRBs are $25$, $50$, and $100$, respectively. We can conclude that the total processing time and computing resources were mainly spent on the modulation, demodulation, coding, and decoding. These tasks played the bigger roles in terms of complexity and runtime overhead in the BBU protocol stack. 

% The maximum traffic load was recorded at $\rm{MCS} = 27$ and $\rm{PRBs} = 100$. 
% The maximum load which was at MCS 27 provided with OAI software has been considered during running  on VMware VM. Also we have depended the worse case of wire channel condition which was 80 dB that can be adjusted by the attenuators between UL and DL links. 

%The results in Figure~\ref{CPU_per} can be fit to the following equation,
%\begin{equation}
%{CP{{U}_{\% }}(x)=\lambda {{x}^{2}}+\rho x+\gamma}
%\end{equation}where $x$ represents the MCS value and the other parameters: $\lambda$, $\rho $, and $\gamma $ are defined in Table~\ref{fit2}. 

%\begin{table}[h!]
%\renewcommand{\arraystretch}{1.5}
%\caption{Parameters}\label{fit2}
%\centering
%\begin{tabular}{|c|c|c|c|}
%\hline
%PRB & $\lambda$ & $\rho$ & $\gamma $\\
%\hline
%25 &  0.1229 & -1.377 & 19.92  \\
%\hline
%50 & 0.1153&  -0.9973   &  24.07 \\
%\hline
%100 &0.1097   &-0.8269  &   31.11 \\
%\hline
%\end{tabular}
%\end{table}

%\subsection{Downlink Maximum Throughput}
To understand better the BBU computational consumption in C-RAN with respect to the users' traffic demand, we will now establish the relationship between the DL throughput and the percentage of CPU usage at the BBU. To begin, we learn that OAI supports $28$ different MCSs with index ranging from $0$ to $27$. In the downlink direction, MCSs with the index $0$ to $9$ are modulated using QPSK, index $10$ to $16$ are modulated using 16-QAM, and the rest are based on 64-QAM. For instance, in LTE FDD system with PRB $100$, corresponding to bandwidth of $20~\rm{MHz}$, we can get $12 \times 7 \times 2 = 168$ symbols per $\rm{ms}$, in case of normal Cyclic Prefix~(CP)~\cite{dahlman20134g}, which is equivalent to a data rate of $16.8~\rm{Mbps}$. Based on the MCS index used in each experiment, we can calculate the corresponding DL throughput by multiplying the bit rate by the number of bits in the modulation scheme. 

\begin{figure}[!ht]
 \centering
 \includegraphics[width=0.4\textwidth]{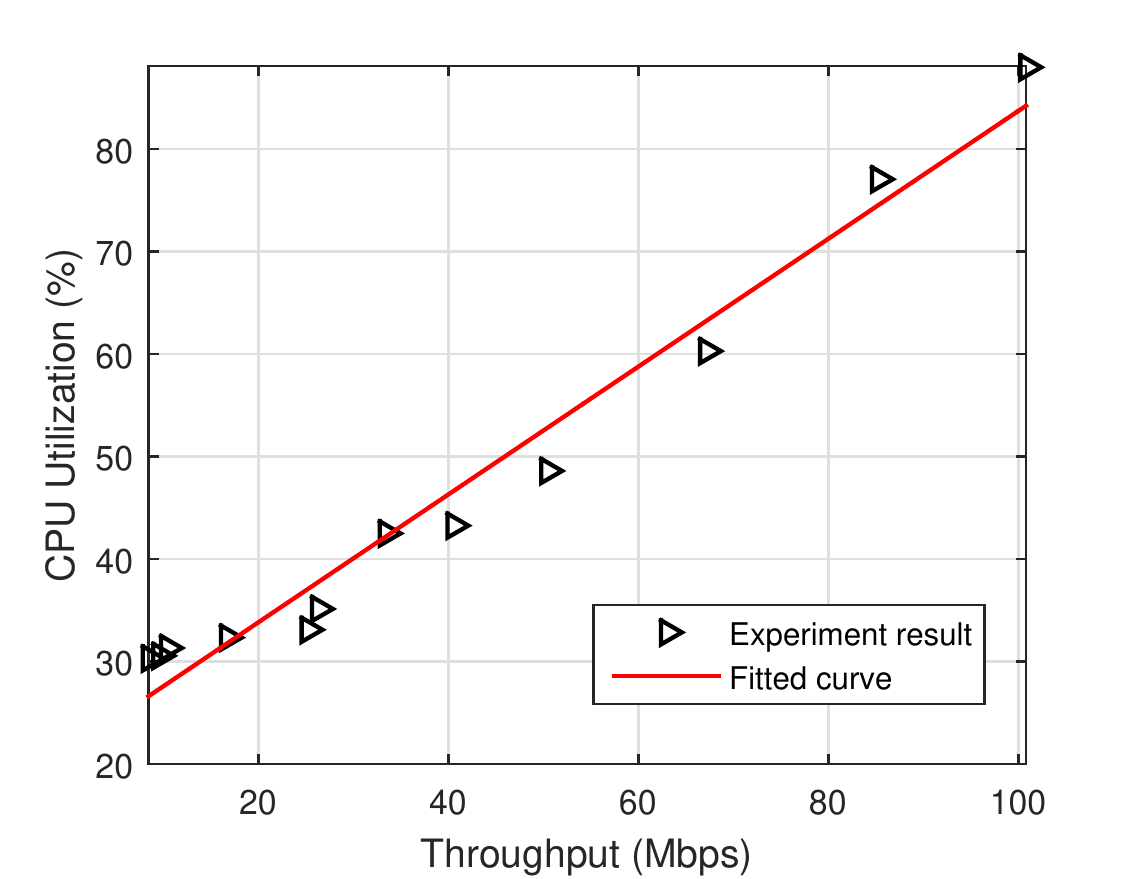}
\caption{Percentage of CPU usage versus the downlink throughput.}\label{line}
\end{figure}

Figure~\ref{line} shows the CPU utilization percentage at the BBU corresponding to different DL throughputs. Using the calculated results, we have fitted the CPU utilization as a linear function of the DL throughput as,
\begin{equation}
{\rm{CPU}}\left[ \%  \right] = 0.6237\phi + 21.3544,
\end{equation} where $\phi$ is the throughput measured in $\rm{Mbps}$. 
%\begin{equation}
%{CP{{U}_{\%}}=a*throupu{{t}_{(Mbps)}}+b}
%\end{equation}
%where $a$ and $b$ parameters equal to 0.6237 and 21.3544, respectively. 

% \section{Future Work} \label{sec:future}
% Our work is a first step towards the realization of a real-time operating system and virtualization environment in C-RAN. In VM, the software radio application runs on a hypervisor and the hardware is either fully or partially virtualized. On the other hand, our testbed can be built in different environments such as Juju~\cite{jujucharmscom}---an open-source platform allowing user to deploy, configure, manage, provision, and dispose OAI eNB over various cloud platforms such as openstack and VMWare; moreover, it can be built in containers (e.g., Linux Container LXC and Docker). This procedure will help upgrade the testbed for various environment and give better understanding of the OAI's performance inside different containers. Another line of investigation is on Infrastructure-as-a-Service~(IaaS), where virtualized BBU resources can be adjusted on-demand by a hypervisor~\cite{sabella2013ran}. In our case, we can create more than one OAI VM working on top of a single physical computer implemented by using a VMware hypervisor that manages several virtual servers. This approach will be a practical solution for testing parallel OAI eNBs on one hypervisor.

%---------------------------------------------------   
\section{Conclusions}\label{sec:conclusion}
To exploit the benefits of C-RAN, software implementation coupled with real-hardware experiments is essential to understand the runtime complexity and performance limits of virtual Baseband Units~(BBUs). We studied and analyzed several aspects related to the practical implementation of the C-RAN architecture to support 5G systems. First, we presented the main C-RAN testbed implementation challenges and studied several virtualization approaches. Second, by using an OAI emulation platform, we built the eNB and UE on low-latency Linux VMs. Experiments were carried out to evaluate the BBU performance under various computing and radio-resource configurations. Our experimental results showed that the frame processing time and CPU utilization of the BBU increase with the PRB resource and MCS index. Third, based on these results, we established empirical models for the estimation of the BBU processing time as a function of CPU frequency, MCS, and PRB index, as well as for the BBU's CPU usage as a linear function of the downlink throughput. These models provide real-world insights into the computational requirements of the BBU, and may serve as key inputs for the design of resource-management solutions in C-RAN systems. 

% In this paper, we present a programmable C-RAN testbed where the BBU is virtualized using the OpenAirInterface (OAI) software platform and the eNodeB and User Equipments (UEs) are implemented using the USRP boards. Extensive experiments have been performed in a FDD downlink LTE emulation system to characterize  the performance and computing resource consumption of the BBU under various conditions. It is shown that the processing time and CPU utilization of the BBU increase with the channel resources and Modulation and Coding Scheme (MCS). Specifically, the CPU utilization percentage can be well approximated as a linear increasing function of the maximum downlink data rate. These results provide real-world insights into the characteristics of the BBU in terms of computing resource and power consumption which serve as critical inputs for design of efficient resource provisioning and allocation in C-RAN systems.

\textbf{Acknowledgments:}
This work was partially supported by the US National Science Foundation Grant No.~CNS-1319945.

\balance

\bibliographystyle{ieeetr}\small
\bibliography{2017_ICAC_CRAN,TWC17}

\end{document}